\documentclass[12pt]{iopart}

\usepackage[T1]{fontenc}
\usepackage[a4paper]{geometry}
\usepackage{float}
\usepackage{units}
\usepackage{textcomp}
  \expandafter\let\csname equation*\endcsname\relax
  \expandafter\let\csname endequation*\endcsname\relax
 \expandafter\let\csname eqnarray\endcsname\relax
  \expandafter\let\csname ende\endcsname\relax
\usepackage{amsmath}
\usepackage{amssymb}
\usepackage{esint}

\usepackage{graphics,graphicx,epsfig}

\begin{document}

\title{Semiclassical matrix elements for a chaotic propagator in the Scar
functions basis}

\author{Alejandro M.F Rivas}

\address{Departamento de F\'{\i}sica, Comisi\'on Nacional de Energ\'{\i}a At\'omica. Av.
del Libertador 8250, 1429 Buenos Aires, Argentina. \\
 }

\thanks{member of the CONICET}

\date{\today {}}
\begin{abstract}
A semiclassical approximation for the matrix elements of a quantum
chaotic propagator in the scar function basis has been derived. The
obtained expression is solely expressed in terms of canonical invariant objects.
For our purpose, we have used, the recently developed, semiclassical
matrix elements of the propagator in coherent states, together with
the linearization of the flux in the neighborhood of a classically
unstable periodic orbit of chaotic two dimensional systems. The expression
here derived is successfully verified to be exact for a (linear) cat
map, after the theory is adapted to a discrete phase space appropriate
to a quantized torus.
\end{abstract}


\maketitle

\section{Introduction}

The Gutzwiller trace formula provides a tool for the semi-classical
evaluation of the energy spectrum of a classically chaotic Hamiltonian
system in terms of canonical invariants of periodic orbits (POs).
However the number of long periodic orbits required to resolve the
spectrum increases exponentially with the Heisenberg time $T_{H}$
\cite{gutz}. For this reason, the approach is limited to eigenenergies
close to the ground state \cite{4ver}. Of course, Gutzwiller's formula
is very attractive because it is given in terms of canonical invariants,
and for this reason a lot of work has been dedicated to improve this
theory \cite{5ver}. 
In particular, the paper by Mehlig and Wilkinson \cite{mehlig}, formulates 
the Guzwiller trace  formula using coherent sates. This manifestly 
canonically invariant formulation allows to separate the contribution of 
each PO to the quantum evolution operator.

The semiclassical theory of short periodic orbits developed by Vergini
and co-workers \cite{6ver}-\cite{13ver} is a formalism where the
number of used POs increases only linearly with the mean energy density,
allowing to obtain all the quantum information of a chaotic Hamiltonian
system in terms of a very small number of short periodic orbits. The
key elements in this theory are wave functions related to short unstable
POs and then, it is crucial the evaluation of matrix elements between
these wave functions.

In this context, these wave functions, named scar function because
of its strong connection with the scarring phenomenon \cite{8ver},
are not restricted to a PO; it additionally includes dynamical information
up to the Ehrenfest time and, as a result, is influenced by pieces
of the stable and unstable manifolds near the PO. These wave functions
define an optimal basis in chaotic systems \cite{6ver,9ver} and then,
they have attracted an increasing interest in closed systems \cite{10ver}.
Moreover, they have been shown to be crucial for the understanding
of long living resonances in open systems \cite{11ver}-\cite{Lisandro2}.

In recent work, estimates for the asymptotic behavior of off-diagonal
matrix elements \cite{edudavid}, and asymptotic expansions for matrix
elements in the quantum cat maps \cite{13ver} were derived. Also,
recently a general semi-classical expression in phase space for the
scar functions was obtained explicitly in terms of the classical invariants
that generates the dynamics of the system \cite{scarSC}. 

In order to perform further developments to this theory of short periodic
orbits, the semiclassical evaluation of matrix elements in a scar
function basis set is an important objective. With these matrix elements
at hand, the energy spectrum can be obtained without requiring an
explicit computation of scar functions. This is the purpose of this
paper. For our objective we use the recently developed semiclassical
matrix elements of the quantum propagator between coherent states
\cite{aleprl}. After what, we perform the needed time integrals to
obtain the matrix elements of the propagator between scar functions. 

Also, for two degrees of freedom Hamiltonian systems the dynamics
is studied entirely within a surface of section transversal to the
periodic orbit. That is, the full dynamics is studied through a two
dimensional section map. With the purpose of showing the validity
of our approximation, we have compared the general expression here
founded with numerical calculations in a \textquotedbl{}realistic\textquotedbl{}
system, the cat map, i.e. the quantization of linear symplectic maps
on the torus. As was shown by Keating \cite{keat2} in this case the
semiclassical theory is exact, making this maps an ideal probe for
our expression. After the formalism is adapted for a torus phase space,
we see that the semiclassical expression here deduced coincide exactly
with the numerically computed matrix elements for the cat maps. 

Of course, in order to include nonlinear contributions a deeper understanding
of the dynamics up to the Ehrenfest time is required; in this respect,
enormous efforts were recently carried out in such a direction for
the diagonal matrix elements \cite{epldiag}.

This paper is organized as follows. In section 2 we introduce the
definition of scar functions in terms of coherent states. Hence, we
review the construction of the semiclassical matrix elements of the
propagator in the coherent states basis \cite{aleprl}. We then show
the utility of introducing the Weyl representation, and that our approach
avoids for any complex trajectory.

Section 3 is devoted to obtain a semi classical expression for the
matrix elements of the propagator in the scar function basis. For
that purpose, we need to perform a linearization of the flux close
to periodic orbits and to express the classical evolutions in the
stable and unstable directions. Also, we obtain the expressions for
a system with a discrete time evolution.

In Section 4 we study the particular case of the cat map where not
only the semi-classical theory is exact but also the linear approximation
is valid throughout the torus. After the semi classical expressions
here deduced are adapted for a torus phase space, we see that they
coincide exactly with the numerically computed matrix elements for
the cat maps.

\section{Coherent states matrix elements}

Scar function states are the main object of study of the current work.
According to \cite{depola}-\cite{faure}, the scar function $\left|\varphi_{X}^{\phi}\right\rangle $
of parameter $\phi$ constructed on a single periodic point $X=\left(P,Q\right)$
is defined as 
\begin{equation}
\left|\varphi_{X}^{\phi}\right\rangle =\int_{-\infty}^{\infty}dte^{i\phi t}f_{T}(t)\widehat{U}^{t}\left|X\right\rangle ,\label{scarfun1}
\end{equation}
where $T=\ln\hbar$ is the Ehrenfest time, $\left|X\right\rangle $
is a coherent state centered in the point $X$ on the periodic orbit
and $\widehat{U}^{t}$ is the unitary propagator that governs the
quantum evolution of the system. While $f_{T}(t)$ is a decaying function
that takes negligible values for $|t|>T/2$. Without loss of generality,
it will be convenient for our purpose to choose $f_{T}(t)=e^{-\left(\frac{4t}{T}\right)^{2}}$,
that is the scar function is 
\begin{equation}
\left|\varphi_{X}^{\phi}\right\rangle =\int_{-\infty}^{\infty}dte^{i\phi t}e^{-\left(\frac{4t}{T}\right)^{2}}\widehat{U}^{t}\left|X\right\rangle .\label{scarfun}
\end{equation}
This wavefunctions have been shown, in the Husimi representation,
to live in the neighborhood of the trajectory, resembling the hyperbolic
structure of the phase space in their immediate vicinity \cite{faure},
while its Wigner function also shows hyperbolic fringes asymptotic
to the stable and unstable manifolds \cite{scarSC}. 
Wigner functions with hyperbolic structure have been spotted in previous works. For instance the pioneer work of Berry \cite{berry} shows this phenomenon for the spectral Wigner function in continuous systems while for maps this has been shown in \cite{SWF}. While in the paper of Nicacio et al. \cite{nicacio} the hyperbolic fringes are observed for a superposition of two squeezed states with orthogonal squeezing directions.
As is mention by Nicacio et  al. \cite{nicacio} the scar functions are superpositions of Gaussian states with different degrees (and directions) of squeezing, i.e., they are generalized Gaussian
cat states.

The purpose of this work, is to study semiclassically the matrix elements
of the the unitary propagator in the scar function basis, 
\[
\left\langle \varphi_{X_{1}}^{\phi_{1}}\right|\widehat{U}^{t}\left|\varphi_{X_{2}}^{\phi_{2}}\right\rangle .
\]
From the definition of scar functions (\ref{scarfun1}) we get, 
\begin{equation}
\left\langle \varphi_{X_{1}}^{\phi_{1}}\right|\widehat{U}^{t}\left|\varphi_{X_{2}}^{\phi_{2}}\right\rangle =\int_{-\infty}^{\infty}\int_{-\infty}^{\infty}dt_{1}dt_{2}e^{i\left(\phi_{2}t_{2}-\phi_{1}t_{1}\right)}f_{T}(t_{1})f_{T}(t_{2})\langle X_{1}|\widehat{U}^{\left(t+t_{2}-t_{1}\right)}|X_{2}\rangle.\label{eq: phiUphi}
\end{equation}
Hence, we need to calculate $\langle X_{1}|\widehat{U}^{t}|X_{2}\rangle$
the matrix elements of the propagator in the coherent states basis.
The semi-classical matrix elements of the propagator in coherent states
basis have been obtained in \cite{aleprl}, we will here reproduce
the main steps of the procedure. 

Let us write the propagator $\widehat{U}^{t}$ in terms of its, symplectically
invariant, center or Weyl Wigner symbol $U^{t}(x)$ \cite{ozrep},
\begin{equation}
\widehat{U}^{t}\!=\!\frac{1}{\left(\pi\hbar\right)^{L}}\!\int\! dxU^{t}(x)\widehat{R}_{x}\!\quad\!\textrm{and }\!\quad\! U^{t}(x)=\! tr\!\left[\widehat{R}_{x}\widehat{U}^{t}\right],\label{eq:SWF}
\end{equation}
where$\int dx$ is an integral over the whole phase space of $L$
degrees of freedom, while $\widehat{R}_{x}$ denotes the set of reflection
operators thought points $x=\left(p,q\right)$ in phase space \cite{ozrep,opetor} 
(see Appendix). Hence the coherent sates matrix elements can be written
in terms of reflexions as

\begin{equation}
\langle X_{1}|\widehat{U}^{t}|X_{2}\rangle=\left(\frac{1}{\pi\hbar}\right)^{L}\int\langle X_{1}|U^{t}(x)\widehat{R}_{x}|X_{2}\rangle dx.\label{eq:XURX}
\end{equation}
The coherent states on points $X=\left(P,Q\right)$ in phase space
are obtained by translating to $X$ the ground state of the harmonic
oscillator, its position representation is 
\begin{equation}
\left\langle q\right.\left|X\right\rangle =\left(\frac{m\omega}{\pi\hbar}\right)^{\frac{1}{4}}\exp\left[-\frac{\omega}{2\hbar}(q-Q)^{2}+i\frac{P}{\hbar}\left(q-\frac{Q}{2}\right)\right].\label{eq:qX}
\end{equation}
 For simplicity, unit frequency $\left(\omega=1\right)$ and mass
$\left(m=1\right)$ are chosen for the harmonic oscillator without
loss of generality. The overlap of two coherent states is then 
\begin{equation}
\langle X\left|X^{\prime}\rangle\right.=\exp\left[-\frac{\left(X-X^{\prime}\right)^{2}}{4\hbar}-\frac{i}{2\hbar}X\wedge X^{\prime}\right],\label{cohcoh}
\end{equation}
 with the wedge product 
\[
X\wedge X^{\prime}=PQ^{\prime}-QP^{\prime}=\left({\mathcal{J}}X\right).X^{\prime}.
\]
 The second equation also defines the symplectic matrix ${\mathcal{J}}$,
that is 
\begin{equation}
{\mathcal{J}}=\left[\begin{array}{cc}
0 & -1\\
1 & 0
\end{array}\right].
\end{equation}
 As is shown in the Appendix the action of the reflection operator
$\widehat{R}_{x}$ on a coherent state $\left|X\right\rangle $ is
the $x$ reflected coherent state 
\begin{equation}
\widehat{R}_{x}\left|X\right\rangle =e^{\frac{i}{\hbar}X\wedge x}\left|2x-X\right\rangle .\label{reflexcoh}
\end{equation}
 Inserting (\ref{reflexcoh}) and (\ref{cohcoh}) in (\ref{eq:XURX})
the propagator in coherent states is obtained from the Weyl propagator

\begin{equation}
\langle X_{1}|\widehat{U}^{t}|X_{2}\rangle=\frac{1}{\left(\pi\hbar\right)^{L}}e^{\frac{iX_{1}\wedge X_{2}}{2\hbar}}\int U^{t}(x)e^{\left[\frac{i}{\hbar}x\wedge\xi_{0}-\frac{\left(\bar{X}-x\right)^{2}}{\hbar}\right]}dx,\label{eq:XUgaussX}
\end{equation}
with $\xi_{0}\equiv\left(X_{1}-X_{2}\right)$ the chord joining the
points $X_{1}$ and $X_{2}$, while $\bar{X}=\nicefrac{1}{2}\left(X_{1}+X_{2}\right)$
denotes their mid point. 

Also, the semi-classical approximation for the propagator in the Weyl
representation was performed in \cite{ozrep} so that 
\begin{equation}
U^{t}(x)_{SC}=\sum_{\gamma}\frac{2^{L}\exp\left\{ i\hbar^{-1}S_{\gamma}^{t}(x)
+i \frac{\pi}{2}\alpha_{\gamma}^{t}\right\} }{|\det(\mathcal{M}_{\gamma}^{t}+1)|{}^{\frac{1}{2}}}.\label{eq:USC-1}
\end{equation}
 where the summation is over all the classical orbits $\gamma$ whose
center lies on the point $x$ after having evolved a time $t$ \cite{ozrep}.
Then $S_{\gamma}^{t}(x)$ is the classical center generating function
of the orbit, from which the chord $\xi$ joining the initial and
final point of the orbit is obtained 
\begin{equation}
\xi=-{\cal J}\frac{\partial S_{\gamma}^{t}(x)}{\partial x}.\label{cuerda}
\end{equation}
 While $\mathcal{M}_{\gamma}^{t}=\frac{\partial^{2}S_{\gamma}^{t}(x)}{\partial x^{2}}$
stand for the monodromy matrix and $\alpha_{\gamma}^{t}$ its \textit{Morse
index}. 

The metaplectic operators form a "double covering" of the symplectic matrices,
since this property gives contributions to the Morse index \cite{mehlig}.
If we follow the evolution of the symplectic matrix as the trajectory
evolves, each time $\mathcal{M}_{\gamma}^{t}$ crosses a manifold 
where $\det(\mathcal{M}_{\gamma}^{t}+1)=0$ (caustic) 
the path contribution undergoes a divergence changing
the sign from $-\infty$ to $\infty$. This change of the
sign lets the quantum phase proceed by $\frac{\pi}{2}$.
The Morse index $\alpha_{\gamma}^{t}$ therefore changes by $\pm 1$ when crossing caustics \cite{berry,gutz2}.

For sufficiently short times such that the variational problem
has an unique solution there will have a single chord. Although for
longer times, there will be bifurcations producing more chords. In
the case of a single orbit, the corresponding \textit{Morse index}
$\alpha_{\gamma}^{t}=0$. 

The semiclassical approximation for the propagator (\ref{eq:USC-1})
is inserted in (\ref{eq:XUgaussX}) so that

\begin{equation}
\langle X_{1}|\widehat{U}_{SC}^{t}|X_{2}\rangle=\left(\frac{2}{\pi\hbar}\right)^{L}e^{\frac{i}{2\hbar}X_{1}\wedge X_{2}}\sum_{\gamma}e^{i\alpha_{\gamma}^{t}}\int\frac{\exp-\frac{1}{\hbar}\left(\bar{X}-x\right)^{2}}{\left|\det\left(\mathcal{M}_{\gamma}^{t}+1\right)\right|^{\frac{1}{2}}}\exp\frac{i}{\hbar}\left[S_{\gamma}^{t}(x)-\xi_{0}\wedge x\right]dx.\label{eq:XUscX}
\end{equation}
 In order to perform the phase space integral in (\ref{eq:XUscX}),
it must be noted that classical orbits that start near $X_{2}$ and
end near $X_{1}$ will have an important contribution in (\ref{eq:XUscX}).
These orbits have their center points close to $\bar{X}$. Hence,
let us expand the center action up to quadratic terms near the mid
point $\bar{X}$, so that, 
\begin{equation}
S_{\gamma}^{t}(x)=S_{\gamma}^{t}(\overline{X})+\overline{\xi}\wedge x^{\prime}+x^{\prime\dagger}\mathit{B}_{t}x^{\prime}+O(x^{\prime3})\label{action}
\end{equation}
 with $x=\overline{X}+x^{\prime}$ and $S_{\gamma}^{t}(\overline{X})$
is the action of the orbit through the point $\overline{X}$ for which
the chord $\overline{\xi}$ is 

\[
\overline{\xi}=\left.-{\cal J}\frac{\partial S_{\gamma}^{t}(x)}{\partial x}\right|_{x=\overline{X}},
\]
 while, the symmetric matrix $B_{t}$ is the Cayley representation
of $\mathcal{M}_{\gamma}^{t}$ 
\begin{equation}
{\mathcal{J}}\mathit{B}_{t}=\frac{1-\mathcal{M}_{\gamma}^{t}}{1+\mathcal{M}_{\gamma}^{t}}=\frac{1}{2}\frac{\partial^{2}S_{\gamma}^{t}(x)}{\partial x^{2}}.\label{eq:JB}
\end{equation}
Let us define the action $\widetilde{S}_{\gamma}^{t}(\overline{X})=S_{\gamma}^{t}(\overline{X})+\hbar\frac{\pi}{2}\alpha_{\gamma}^{t}$
in order to include the Morse index in the action. After the linearization
of the flux around the middle point, expression (\ref{action}) is
inserted in (\ref{eq:XUscX}) hence we get
\begin{equation}
\langle X_{1}|\widehat{U}_{SC}^{t}|X_{2}\rangle=\left(\frac{2}{\pi\hbar}\right)^{L}\sum_{\gamma}\frac{\exp\frac{i}{\hbar}\left[\widetilde{S}_{\gamma}^{t}(\overline{X})-\xi_{0}\wedge\overline{X}+\frac{1}{2}X_{1}\wedge X_{2}\right]}{\left|\det\left(\mathcal{M}_{\gamma}^{t}+1\right)\right|^{\frac{1}{2}}}\times I,\label{eq:UI}
\end{equation}
 with 
\begin{equation}
I=\int\exp\frac{1}{\hbar}\left[-x^{\prime\dagger}\mathrm{\mathbf{\mathit{C}}}x^{\prime}+i\left[x^{\prime\dagger}\mathit{B}_{t}x^{\prime}+\left(\overline{\xi}-\xi_{0}\right)\wedge x^{\prime}\right]\right]dx^{\prime}
\end{equation}
a quadratic integral. The matrix$\mathbf{\mathrm{\mathbf{\mathit{C}}}}$
is the quadratic form that denotes the scalar product, 
\[
x^{\prime2}=x^{\prime}.x^{\prime}=x^{\prime\dagger}\mathrm{\mathbf{\mathit{C}}}x^{\prime},
\]
where $x^{\dagger}$ denotes the transposed vector. We now perform
exactly the quadratic integral, using

\begin{equation}
I=\int\exp\left[-\frac{1}{\hbar}x^{\prime\dagger}\mathcal{V}_{t}x^{\prime}+\frac{1}{\hbar}Y.x^{\prime}\right]dx^{\prime}=\frac{\left(\pi\hbar\right)^{L}}{\sqrt{\left(\det\mathcal{V}_{t}\right)}}\exp\left[\frac{1}{4\hbar}Y^{\dagger}\mathcal{V}_{t}^{-1}Y\right].\label{eq:IYY}
\end{equation}
 From equation (\ref{eq:UI}) 
\begin{equation}
\mathcal{V}_{t}=\mathbf{\mathit{C}}-i\mathit{B}_{t}\label{eq: VCB}
\end{equation}
 and 
\begin{equation}
Y=i{\cal J\mathrm{\left(\overline{\xi}-\xi_{0}\right)=2i{\cal J\mathrm{\delta_{t}}}}},\label{eq:Ydelta}
\end{equation}
 where $\overline{\xi}=x_{f}-x_{i}$ is the chord that joins $x_{f}$
and $x_{i}$, respectively the final and initial point of the orbit
of center $\overline{X}$. This last expression defines the point
shift $\delta_{t}$, so that 
\begin{equation}
\delta_{t}=\frac{1}{2}\left(\overline{\xi}-\xi_{0}\right)=x_{f}-X_{1}=X_{2}-x_{i},\label{eq:deltadef}
\end{equation}
Note that, the point shift $\delta_{t}$ is zero if there is a classical
orbit starting in the point $X_{2}$ and ending in $X_{1}$. Inserting
(\ref{eq:IYY}) in (\ref{eq:UI}), we get for the propagator in coherent
states, 
\begin{equation}
\langle X_{1}|\widehat{U}_{SC}^{t}|X_{2}\rangle=2^{L}\sum_{\gamma}\frac{\exp\frac{i}{\hbar}\left[\widetilde{S}_{\gamma}^{t}(\overline{X})-\frac{1}{2}X_{1}\wedge X_{2}\right]}{\left[\det\mathcal{V}_{t}\left|\det\left(\mathcal{M}_{\gamma}^{t}+1\right)\right|\right]^{\frac{1}{2}}}\times\exp\left[\frac{-1}{\hbar}\delta_{t}^{\dagger}\mathcal{\widetilde{V}}\delta_{t}\right],\label{eq:UYY}
\end{equation}
with the complex matrix $\mathcal{V}_{t}$ and the point shift $\delta_{t}$
defined respectively in (\ref{eq: VCB}) and (\ref{eq:deltadef})
while $\mathcal{\widetilde{V}}={\mathcal{J}^{\dagger}}\mathcal{V}_{t}^{-1}{\mathcal{J}}$. 

In order to separate amplitude and phase terms in (\ref{eq:UYY})
, it is useful to write

\begin{equation}
\mathcal{\widetilde{V}}={\mathcal{J}^{\dagger}}\mathcal{V}_{t}^{-1}{\mathcal{J}}={\mathcal{J}^{\dagger}}\frac{1}{\mathit{C}-i\mathit{B}_{t}}{\mathcal{J}}=\mathcal{\overline{C}}_{t}-i\mathcal{\overline{B}}_{t},\label{eq:VCB1}
\end{equation}
with the real matrices 
\[
\mathcal{\overline{C}}_{t}=\Re(\mathcal{\widetilde{V}})\quad\textrm{and}\quad\mathcal{\overline{B}}_{t}=-\Im(\mathcal{\widetilde{V}}).
\]
Also, 
\begin{equation}
\det\mathcal{V}_{t}=\biggl|\det\mathcal{V}_{t}\biggr|e^{i\varepsilon},\label{eq:detV}
\end{equation}
with $\biggl|\det\mathcal{V}_{t}\biggr|$ denoting the modulus and
$\varepsilon$ the argument.

Hence inserting (\ref{eq:VCB1}) and (\ref{eq:detV}) in the~matrix
elements of the coherent state propagator (\ref{eq:UYY}) we obtain

\begin{align}
\langle X_{1}|\widehat{U}_{SC}^{t}|X_{2}\rangle & = & 2^{L}\sum_{\gamma}\frac{1}{\left[\left|\det\mathcal{V}_{t}\right|\left|\det\left(\mathcal{M}_{\gamma}^{t}+1\right)\right|\right]^{\frac{1}{2}}}\exp\left[-\frac{\delta_{t}^{\dagger}\mathcal{\overline{C}}_{t}\delta_{t}}{\hbar}\right]\nonumber \\
 & \times & \exp\frac{i}{\hbar}\left[\widetilde{S}_{\gamma}^{t}(\overline{X})-\frac{1}{2}X_{1}\wedge X_{2}+\delta_{t}^{\dagger}\mathcal{\overline{B}}_{t}\delta_{t}+\hbar\frac{\varepsilon}{2}\right].\label{eq: XUscUdiag-1}
\end{align}
 This last expression of the semiclassical matrix elements between
two coherent states of the quantum propagator is entirely expressed
in terms of real classical objects, namely the action $S_{\gamma}^{t}(\bar{X})$
of the classical real orbit whose mid point is $\bar{X}$, the point
shift $\delta_{t}$, the monodromy matrix ${\cal M}_{\gamma}$ and
its Cayley representation $B_{t}$ and $\mathit{C}$, the scalar product
form. We must note that the phase term in the second line of the expression
is clearly separated from the amplitude ones, in the first line. In
this way it is important to remark the Gaussian term that dampens
the amplitude for large values of the point shift $\delta_{t}$, that
is for orbits centered on $\overline{X}$ that start far from the
point $X_{2}$ (then end far from $X_{1}$). So as, the main contribution
in the sum over classical orbits in (\ref{eq: XUscUdiag-1}) will
come from the particular orbit $\gamma$, centered in $\overline{X}$
, whose initial point $x_{i}$ lies the closest to $X_{2}$. Other
orbits contributions will be highly damped by the exponential term
involving the point shift $\delta_{t}$. Then, only this particular
orbit will be taken into account in the next sections in order to calculate
 the matrix elements for scar functions. 

It must be mentioned here that, extensive work has been previously performed for the 
coherent states matrix elements of the propagator. In particular, a complete semiclassical derivation was performed by Baranger et al. \cite{baranger}, while dos Santos and de Aguiar performed a Weyl ordering treatment in \cite{aguiar}.
Although mathematically correct, both constructions involve an analytic continuation to complex trajectories, while expression (\ref{eq: XUscUdiag-1}),
derived originally in \cite{aleprl}, has the peculiarity to avoid
complex trajectories, only the real canonical variables of the classical system are needed.

Also, note that, if $t=0$, the quantum propagator
is just the identity operator in Hilbert space, the classical symplectic
matrix is the identity, the center action is null, and so are the
symmetric matrix $\left(\mathit{B}_{t=0}=0\right)$ and the chord
$\overline{\xi}=2\delta_{t}-X_{2}+X_{1}=0$ . Hence 
\begin{align*}
\langle X_{1}|\widehat{U}_{SC}^{0}|X_{2}\rangle & = & \langle X_{1}|X_{2}\rangle=2^{L}\frac{\exp\frac{i}{\hbar}\left[-\frac{1}{2}X_{1}\wedge X_{2}\right]}{\left|\det\left(2\right)\right|^{\frac{1}{2}}}\times\exp\left[-\frac{1}{4\hbar}\left(X_{2}-X_{1}\right)^{2}\right]\\
 & = & \exp\left[-\frac{\left(X_{2}-X_{1}\right)^{2}}{4\hbar}-\frac{i}{2\hbar}X_{1}\wedge X_{2}\right],
\end{align*}
 and we recover the result (\ref{cohcoh}) for the overlap of coherent
sates. As we have seen in \cite{aleprl} expression (\ref{eq: XUscUdiag-1})
is exact for the case of linear systems.

\section{Matrix elements for Scar functions}

For the study of the matrix elements of the quantum propagator in
the scar function basis, we must insert the expression for the matrix
elements of the propagator in coherent sates basis (\ref{eq: XUscUdiag-1})
in the scar function matrix elements expression (\ref{eq: phiUphi}).
In that way we get,

\begin{align}
\left\langle \varphi_{X_{1}}^{\phi_{1}}\right|\widehat{U}_{SC}^{t}\left|\varphi_{X_{2}}^{\phi_{2}}\right\rangle  & = & 2^{L}\exp\frac{i}{\hbar}\left[-\frac{1}{2}X_{1}\wedge X_{2}\right]\int_{-\infty}^{\infty}\int_{-\infty}^{\infty}dt_{1}dt_{2}e^{i\left(\phi_{2}t_{2}-\phi_{1}t_{1}\right)}f_{T}(t_{1})f_{T}(t_{2})\nonumber \\
 & \times & \frac{\exp\frac{i}{\hbar}\left[\widetilde{S}_{\gamma}^{t_{R}}(\overline{X})+\delta_{t_{R}}^{\dagger}\mathbf{\mathcal{\overline{B}}}_{t_{R}}\delta_{t_{R}}+\hbar\frac{\varepsilon}{2}\right]}{\left[\left|\det\mathcal{V}_{t_{R}}\right|\left|\det\left(\mathcal{M}_{\gamma}^{t_{R}}+1\right)\right|\right]^{\frac{1}{2}}}\times\exp\left[-\frac{\delta_{t_{R}}^{\dagger}\mathcal{\overline{C}}_{t_{R}}\delta_{t_{R}}}{\hbar}\right],\label{eq:phiUphiSC-4}
\end{align}
 where $t_{R}=t+t_{2}-t_{1}$. Note that, in (\ref{eq:phiUphiSC-4})
only the contribution of a unique orbit is taken into account. As
we have already mentioned the contributions of orbits with longer
point shifts were neglected. With the choice for $f_{T}(t)$ made
in (\ref{scarfun}) and performing the change of variables $t_{s}=t_{2}+t_{1}$
and $t_{r}=t_{1}-t_{2}$ it is possible to separate part of the time
integrals so that,

\begin{align}
\noindent\left\langle \varphi_{X_{1}}^{\phi_{1}}\right|\widehat{U}_{SC}^{t}\left|\varphi_{X_{2}}^{\phi_{2}}\right\rangle  & = & \exp\frac{i}{\hbar}\left[-\frac{1}{2}X_{1}\wedge X_{2}\right]\int_{-\infty}^{\infty}dt_{s}e^{\frac{i}{2}\left(\phi_{2}-\phi_{1}\right)t_{s}}e^{-2t_{s}^{2}/T^{2}}\int_{-\infty}^{\infty}dt_{r}e^{-2t_{r}^{2}/T^{2}}e^{-\frac{i}{2}\left(\phi_{2}+\phi_{1}\right)t_{r}}\nonumber \\
 & \times & \frac{\exp\frac{i}{\hbar}\left[\widetilde{S}_{\gamma}^{t-t_{r}}(\overline{X})+\delta_{t-t_{r}}^{\dagger}\mathbf{\mathcal{\overline{B}}}_{t-t_{r}}\delta_{t-t_{r}}+\hbar\frac{\varepsilon}{2}\right]}{\left[\left|\det\mathcal{V}_{t-t_{r}}\right|\left|\det\left(\mathcal{M}_{\gamma}^{t-t_{r}}+1\right)\right|\right]^{\frac{1}{2}}}\times\exp\left[-\frac{\delta_{t-t_{r}}^{\dagger}\mathcal{\overline{C}}_{t-t_{r}}\delta_{t-t_{r}}}{\hbar}\right].\label{eq:phiUphiSC-3-1-1-2}
\end{align}
Performing the first time integral, defining $A=T\sqrt{\frac{\pi}{2}}e^{-\frac{T^{2}\left(\phi_{2}-\phi_{1}\right)^{2}}{32}}$,
and changing the variables to $t'=t-t_{r}$ we get: 
\begin{align}
\left\langle \varphi_{X_{1}}^{\phi_{1}}\right|\widehat{U}_{SC}^{t}\left|\varphi_{X_{2}}^{\phi_{2}}\right\rangle  & = & A\exp\frac{i}{\hbar}\left[-\frac{1}{2}X_{1}\wedge X_{2}\right]\int_{-\infty}^{\infty}dt'e^{-2(t-t'){}^{2}/T^{2}}e^{-\frac{i}{2}\left(\phi_{2}+\phi_{1}\right)(t-t')}\nonumber \\
 & \times & \frac{\exp\frac{i}{\hbar}\left[\widetilde{S}_{\gamma}^{t'}(\overline{X})+\delta_{t'}^{\dagger}\mathbf{\mathcal{\overline{B}}}_{t'}\delta_{t'}+\hbar\frac{\varepsilon}{2}\right]}{\left[\left|\det\mathcal{V}_{t'}\right|\left|\det\left(\mathcal{M}_{\gamma}^{t'}+1\right)\right|\right]^{\frac{1}{2}}}\times\exp\left[-\frac{\delta_{t'}^{\dagger}\mathcal{\overline{C}}_{t'}\delta_{t'}}{\hbar}\right].\label{eqPHIUPHISC}
\end{align}

Equation (\ref{eqPHIUPHISC}) expresses the semicalssical approximation
of the matrix elements of the quantum propagator in the scar function
basis uniquely in terms of classical objects. However, at this point
expression (\ref{eqPHIUPHISC}) has the inconvenient that we need
the evaluation of each one of this classical objects, for all the
times of integration involved. In what follows, we will obtain explicit
expression for the classical objects involved in (\ref{eqPHIUPHISC}).

For that purpose, we will first perform the study on a surface of
section that is transversal to the flux and passing through $\overline{X}$.
In analogy with classical Poincar\'e surfaces of section. The flux restricted
to this section is now a map on the section, for this map the time
is discrete and time integrals must be replaced by summations.

The study of autonomous fluxes through a map on surface of section
is a standard procedure, in the case of billiards this is done through
the well known Birkhoff coordinates. Also, quantum surface of section
methods are shown to be exact \cite{prosen} for general Hamiltonian
systems.

For this procedure, we can choose coordinates near the periodic orbit
of period $\tau$ where $X$$_{2}$ belongs, such that one coordinate
is the energy $E$ and the conjugate coordinate is the time along
the orbit. With this choice of coordinates, a point $x=(\tilde{x},t,E)$
with now $\tilde{x}$ a $(2L-2)$ vector on the so called central
surface of section \cite{ozrep}. In order to perform our study on
this surface of section near the fixed point $X$$_{2}$ we have linearized
the flux in the the neighborhood of the orbit through $X_{2}$. That
is, for the orbit $\gamma$ that starts in $x_{i}$ and end in $x_{f}$:
$x_{f}={\cal L}_{\gamma}^{t}(x_{i})\approx\mathcal{M}_{\gamma}^{t}x_{i}$.
Where $\mathcal{M}_{\gamma}^{t}$ is the symplectic matrix denoting
this linearized time evolution. As was shown in \cite{ozrep}, in
the transformation $x_{f}=\mathcal{M}_{\gamma}^{t}x_{i}$ for times
$t$ that are integer multiples of $\tau$, $t=n\tau$, the points
$x_{f}=(\tilde{x}_{f},t_{f},E_{f})$ and $x_{i}=(\tilde{x}_{i},t_{i},E_{i})$
on the surface of section have the same energy ($E_{f}=E_{i}$) and
time along the orbit ($t_{f}=t_{i}$) so we can write , 
\begin{equation}
\mathcal{M}_{\gamma}^{t}=\left(\begin{array}{c|c}
m_{\gamma}^{t} & 0\\
\hline 0 & 10\\
 & 01
\end{array}\right)
\end{equation}
 with 
\begin{equation}
\det[1+\mathcal{M}_{\gamma}^{t_{1}}]=4\det[1+m_{\gamma}^{t_{1}}]\ ,
\end{equation}
 where $m_{\gamma}^{t_{1}}$ is now the $(2L-2)\times(2L-2)$ symplectic
matrix for the center map determined by the orbit $\gamma$ on the
surface section, that is 
\[
\tilde{x}_{f}=m_{\gamma}^{t}\tilde{x_{i}}
\]
 From now on, the $2L$ dimensional autonomous flux is studied through
the $2L-2$ map on the mentioned surface of section. Also the point
$X_{2}$ on the periodic orbit of the flux is a fixed point for the
map on the section.

For values of the time that are integer multiples of $\tau$ and for
points $\tilde{x}$ on the surface of section, the expression (\ref{eqPHIUPHISC})
for the matrix elements of the propagator takes now the following
form (replacing time integrations by summations):

\begin{align}
\left\langle \varphi_{X_{1}}^{\phi_{1}}\right|\widehat{U}^{t}\left|\varphi_{X_{2}}^{\phi_{2}}\right\rangle  & = & \exp\frac{i}{\hbar}\left[-\frac{1}{2}X_{1}\wedge X_{2}\right]A\sum_{n'=-\infty}^{\infty}e^{-2(t-t'){}^{2}/T^{2}}e^{-\frac{i}{2}\left(\phi_{2}+\phi_{1}\right)(t-t')}\nonumber \\
 & \times & \frac{\exp\frac{i}{\hbar}\left[\widetilde{S}_{\gamma}^{t'}(\overline{X})+\delta_{t'}^{\dagger}\mathbf{\mathcal{\overline{B}}}_{t'}\delta_{t'}+\hbar\frac{\varepsilon}{2}\right]}{\left[\left|\det\mathcal{V}_{t'}\right|\left|\det\left(\mathcal{M}_{\gamma}^{t'}+1\right)\right|\right]^{\frac{1}{2}}}\times\exp\left[-\frac{\delta_{t'}^{\dagger}\mathcal{\overline{C}}_{t'}\delta_{t'}}{\hbar}\right].\label{eq:phiUphiSCmapa-1}
\end{align}
 with $t=n\tau$ and $t'=n'\tau$ with $n$ and $n'$ integer numbers.
This last expression represents the semiclassical propagator matrix
elements in scar function basis, on the surface of section that cuts
transversally the periodic orbit on the point $X_{2}$. In order to
deal with the infinite time summation in (\ref{eq:phiUphiSCmapa-1})
we perform a cut off for values of $|t-t'|$ greater that $T$, the
Ehrenfest time, beyond which the time dependent Gaussian became negligible.
Remember also the discussion according to the choice of the the function
$f_{T}(t)$ in (\ref{scarfun1}).

As we have already mentioned, we need to evaluate the classical objects
involved in (\ref{eq:phiUphiSCmapa-1}) to perform the time summation.
Let us first, obtain expressions for the point shifts $\delta_{t'}$
and the center action $\widetilde{S}_{\gamma}^{t}(\overline{X})$
in terms of the monodromy matrix. For that purpose, we linearize the
flux in the the neighborhood of the fixed point $X_{2}$. That is,
\[
x_{f}=X_{1}+\delta_{t}={\cal L}_{\gamma}^{t}(x_{i})={\cal L}_{\gamma}^{t}(X_{2}-\delta_{t})\approx\mathcal{M}_{\gamma}^{t}(X_{2})-\mathcal{M}_{\gamma}^{t}\delta_{t}=X_{2}-\mathcal{M}_{\gamma}^{t}\delta_{t},
\]
where $\mathcal{M}_{\gamma}^{t}$ is the symplectic matrix denoting
this linearized time evolution in the neighborhood of $X_{2}$ , the
last equality hold because $X_{2}$ is a fixed point. Resolving for
$\delta_{t}$ we get, 
\[
\delta_{t}=-\frac{1}{\mathcal{M}_{\gamma}^{t}+1}\left(X_{1}-X_{2}\right).
\]
 Equivalently we can perform the linearization using the center generating
function near the point $X_{2}$ , so that 
\begin{equation}
S_{\gamma}^{t}(x)=S_{\gamma}^{t}(X_{2})+\left(x-X_{2}\right)^{\dagger}\mathit{B}_{t}\left(x-X_{2}\right)+O(x^{\prime3}).\label{action-1}
\end{equation}
 Note that, the linear term $\xi_{2}\wedge\left(x-X_{2}\right)$ is
not present here since $X_{2}$ is a fixed point, hence the chord
$\xi_{2}$ passing thorough it is null, $\xi_{2}=-{\cal J}\frac{\partial S_{\gamma}^{t}(x)}{\partial x}\biggr|_{X_{2}}$$=0$.
Also, for $X_{2}$ being a fixed point 
\[
S_{\gamma}^{t}(X_{2})=tS_{X_{2}},
\]
where $S_{X_{2}}$is the action of the periodic orbit in $X_{2}$
which the Morse index $\alpha_{\gamma}^{t}=t\alpha_{\gamma}$. Let
us define the action $\widetilde{S}_{X_{2}}=S_{X_{2}}+\hbar \frac{\pi}{2} \alpha_{\gamma}$
in order to include the Morse index in the action.

The chord $\overline{\xi}$ of the orbit $\gamma$ centered in $\overline{X}$
is obtained by performing the derivative of the center generating
function (\ref{action-1}) , 
\[
\overline{\xi}=-{\cal J}\frac{\partial S_{\gamma}^{t}(x)}{\partial x}\biggr|_{\overline{X}}=-2{\cal J}\mathit{B}_{t}\left(\overline{X}-X_{2}\right)=-{\cal J}\mathit{B}_{t}\xi_{0}.
\]
So that (as was already seen): 
\begin{equation}
\delta_{t}=\frac{1}{2}\left(\overline{\xi}-\xi_{0}\right)=-\frac{1}{2}\left({\cal J}\mathit{B}_{t}+1\right)\xi_{0}=-\frac{1}{\mathcal{M}_{\gamma}^{t}+1}\xi_{0}.\label{eq:deltaXX}
\end{equation}
 Hence, the center generating function in the middle point$\overline{X}$
is 

\begin{equation}
\widetilde{S}_{\gamma}^{t}(\overline{X})=t\widetilde{S}_{X_{2}}+\left(\overline{X}-X_{2}\right)^{\dagger}\mathit{B}_{t}\left(\overline{X}-X_{2}\right)=t\widetilde{S}_{X_{2}}+\frac{1}{4}\xi_{0}^{\dagger}\mathit{B}_{t}\xi_{0}.\label{eq:SX}
\end{equation}
It is important to mention that the summation to be performed in (\ref{eq:phiUphiSCmapa-1})
is a summation on the orbits $\gamma$ that starts near $X_{2}$ and
after a time $t$ end up near $X_{1}$, having $\bar{X}$ as center
point, this is a sum on the family of heteroclinic orbits as has been
already seen in \cite{edudavid}. Inserting the expressions (\ref{eq:deltaXX})
and (\ref{eq:SX}) respectively for the point shifts $\delta_{t}$
and the center action $\widetilde{S}_{\gamma}^{t}(\overline{X})$
in the scar function expressions (\ref{eqPHIUPHISC}), we obtain 

\begin{align}
\left\langle \varphi_{X_{1}}^{\phi_{1}}\right|\widehat{U}_{SC}^{t}\left|\varphi_{X_{2}}^{\phi_{2}}\right\rangle  & = & A\exp\frac{i}{\hbar}\left[-\frac{1}{2}X_{1}\wedge X_{2}\right]\sum_{n'=-\infty}^{\infty}e^{-2(t-t'){}^{2}/T^{2}}e^{-\frac{i}{2}\left(\phi_{2}+\phi_{1}\right)(t-t')}\nonumber \\
 & \times & \frac{\exp\frac{i}{\hbar}\left[t'\widetilde{S}_{X_{2}}+\xi_{0}^{\dagger}\left(\frac{1}{4}\mathit{B}_{t'}+D_{t'}\right)\xi_{0}+\hbar\frac{\varepsilon}{2}\right]}{\left[\left|\det\mathcal{V}_{t'}\right|\left|\det\left(\mathcal{M}_{\gamma}^{t'}+1\right)\right|\right]^{\frac{1}{2}}}\exp\left[-\frac{\xi_{0}^{\dagger}E_{t'}\xi_{0}}{\hbar}\right],\label{eq:phiUphiSCcont}
\end{align}
with the matrices $E_{t}$ and $D_{t}$ so that 
\begin{equation}
E_{t}=\left(\frac{1}{\mathcal{M}_{\gamma}^{t}+1}\right)^{\dagger}\mathcal{\overline{C}}_{t}\left(\frac{1}{\mathcal{M}_{\gamma}^{t}+1}\right)\quad\textrm{and }\quad D_{t}=\left(\frac{1}{\mathcal{M}_{\gamma}^{t}+1}\right)^{\dagger}\mathcal{\overline{B}}_{t}\left(\frac{1}{\mathcal{M}_{\gamma}^{t}+1}\right).\label{eq:DyE}
\end{equation}
Equation (\ref{eq:phiUphiSCcont}) is a general expression only in
term of classical objects, its difference from (\ref{eq:phiUphiSCmapa-1})
is that we have used the linearization around the periodic orbit in
order to express both the point shift and the center generating function
only in terms of the monodromy matrix of the linearized transformation.
In this way, the semiclassical approximation of the scar function's
matrix elements involves uniquely the action of the classical orbit
$\widetilde{S}_{X_{2}}$, the scalar product $C$ for the symplectic
basis of vectors and the monodromy matrices $\mathcal{M}_{\gamma}^{t}$
. From this former, we obtain it Cayley representation $\mathit{B}_{t}$
through equation (\ref{eq:JB}), after what the complex matrix $\mathcal{V}_{t}$
is obtained with (\ref{eq: VCB}) and (\ref{eq:detV}) expresses its
exponential form, while the real matrices $\mathcal{\overline{C}}_{t}$
and $\mathcal{\overline{B}}_{t}$ defined in (\ref{eq:VCB1}) allows
to obtain $D_{t}$ and $E_{t}$ though (\ref{eq:DyE}).

Although, in order to perform the time summation we need the classical
objects for all the different times involved. As we will show, for
that purpose, it will be convenient to express them in the basis of
eigenvectors of the symplectic matrix. For the case of a map with
one degree of freedom (corresponding to a two degrees of freedom flux),
this is the stable and unstable vector basis $\left(\vec{\zeta_{u}},\vec{\zeta_{s}}\right)$
where the eigenvalues of the symplectic matrix $\mathcal{M}_{\gamma}^{t}$
are $\exp(-\lambda t)$ and $\exp(\lambda t)$, ($\lambda$ is the
stability or Lyapunov exponent of the orbit). 

Let us then define $x_{s}$ and $x_{u}$ as canonical coordinates
along the stable and unstable directions respectively such that $x=(x_{u},x_{s})=x_{u}\vec{\zeta_{u}}+x_{s}\vec{\zeta_{s}}$
with $\vec{\zeta_{u}}\wedge\vec{\zeta_{s}}=1$. As the basis formed
by $\left(\vec{\zeta_{u}},\vec{\zeta_{s}}\right)$ is non orthonormal,
the scalar product of two vectors takes the form,

\[
x_{1}.x_{2}=x_{1}^{\dagger}\mathrm{\mathit{C}}x_{2}=\left[\zeta_{u}^{2}x_{1u}x_{2u}+\zeta_{s}^{2}x_{1s}x_{1s}+\vec{\zeta_{u}}.\vec{\zeta_{s}}\left(x_{1u}x_{2s}+x_{1s}x_{2u}\right)\right].
\]
 That is, the scalar product matrix is,

\begin{equation}
\mathrm{\mathit{C}}=\left[\begin{array}{cc}\zeta_{u}^{2} & \vec{\zeta_{u}}.\vec{\zeta_{s}}\\
\vec{\zeta_{u}}.\vec{\zeta_{s}} & \zeta_{s}^{2}
\end{array}\right]\label{X-xr}
\end{equation}
 with $\zeta_{u}^{2}=\vec{\zeta_{u}}.\vec{\zeta_{u}}$ and $\zeta_{s}^{2}=\vec{\zeta_{s}}.\vec{\zeta_{s}}$.
Since the transformation from the orthonormal basis $\left(\vec{i},\vec{j}\right)$
to the basis $\left(\vec{\zeta_{u}},\vec{\zeta_{s}}\right)$ is symplectic

\[
\det\mathit{C}=\zeta_{u}^{2}\zeta_{s}^{2}-\left(\vec{\zeta_{u}}.\vec{\zeta_{s}}\right)^{2}=1.
\]
Also, in the $\left(\vec{\zeta_{u}},\vec{\zeta_{s}}\right)$ basis,
\begin{equation}
\mathcal{M}_{\gamma}^{t}+1=2\cosh\left(\frac{\lambda t}{2}\right)\left[\begin{array}{cc}e^{t\nicefrac{\lambda}{2}} & 0\\
0 & e^{-t\nicefrac{\lambda}{2}}
\end{array}\right],\label{eq:M+1}
\end{equation}
hence 
\begin{equation}
\left|\det\left(\mathcal{M}_{\gamma}^{t}+1\right)\right|=4\cosh^{2}\left(\frac{\lambda t}{2}\right),\label{detM}
\end{equation}
is easily obtained only in terms of $\lambda$ and $t$. Analogously, 

\[
\mathcal{M}_{\gamma}^{t}-1=2\sinh\left(\frac{\lambda t}{2}\right)\left[\begin{array}{cc}e^{t\nicefrac{\lambda}{2}} & 0\\
0 & -e^{-t\nicefrac{\lambda}{2}}
\end{array}\right],
\]
while, $\mathit{B}_{t}$, the Cayley parametrization of $\mathcal{M}_{\gamma}^{t}$,
is in this basis 
\begin{equation}
\mathit{B}_{t}=\left[\begin{array}{cc}
0 & \tanh\left(t\lambda/2\right)\\
\tanh\left(t\lambda/2\right) & 0
\end{array}\right].\label{cayley}
\end{equation}
Hence, using the expression of the symmetric matrix $\mathit{B}_{t}$
(\ref{cayley}) and the scalar product (\ref{X-xr}) we get the complex
matrix 
\begin{equation}
\mathcal{V}_{t}=\mathit{C}-i\mathit{B}_{t}=\left[\begin{array}{cc}\zeta_{u}^{2} & \vec{\zeta_{u}}.\vec{\zeta_{s}}-i\tanh\left(t\lambda/2\right)\\
\vec{\zeta_{u}}.\vec{\zeta_{s}}-i\tanh\left(t\lambda/2\right) & \zeta_{s}^{2}
\end{array}\right].\label{eq:Vlanda}
\end{equation}
For which the complex determinant 
\begin{equation}
\det\mathcal{V}_{t}=\left[1+\tanh^{2}\left(t\lambda/2\right)+2i\vec{\zeta_{u}}.\vec{\zeta_{s}}\tanh\left(t\lambda/2\right)\right],\label{eq:detv}
\end{equation}
with modulus

\begin{equation}
\left|\det\mathcal{V}_{t}\right|=\sqrt{\left(1+\tanh^{2}\left(t\lambda/2\right)\right)^{2}+\left(2\vec{\zeta_{u}}.\vec{\zeta_{s}}\tanh\left(t\lambda/2\right)\right)^{2}}\label{eq:moddetv}
\end{equation}
 and argument 
\begin{equation}
\epsilon=\arctan\frac{2\vec{\zeta_{u}}.\vec{\zeta_{s}}\tanh\left(t\lambda/2\right)}{1+\tanh^{2}\left(t\lambda/2\right)},
\end{equation}
can be explicitly written in terms of the time and the Lyapunov exponent.
Now, inverting the matrix $\mathcal{V}_{t}$ (\ref{eq:Vlanda}) we
get, 
\[
\mathcal{V}_{t}^{-1}=\frac{1}{\det\mathcal{V}_{t}}\left[\begin{array}{cc}\zeta_{s}^{2} & -\vec{\zeta_{u}}.\vec{\zeta_{s}}+i\tanh\left(t\lambda/2\right)\\
-\vec{\zeta_{u}}.\vec{\zeta_{s}}+i\tanh\left(t\lambda/2\right) & \zeta_{u}^{2}
\end{array}\right]=\frac{1}{\det\mathcal{V}_{t}}\left(\mathit{C}^{-1}+i\mathit{B}_{t}\right).
\]
 Also, we must note that since the matrix $\mathcal{V}_{t}$ is symmetric,
we get that

\begin{equation}
\mathcal{\widetilde{V}}={\mathcal{J}^{\dagger}}\mathcal{V}_{t}^{-1}{\mathcal{J}}=\frac{\mathcal{V}_{t}}{\det\mathcal{V}_{t}}=\frac{1}{\left|\det\mathcal{V}_{t}\right|^{2}}\left(\Re(\mathbf{\det\mathcal{V}}_{t})-i\Im(\mathbf{\det\mathcal{V}}_{t})\right)\left(\mathit{C}-i\mathit{B}_{t}\right)=\mathcal{\overline{C}}_{t}-i\mathcal{\overline{B}}_{t}.\label{eq:VCB}
\end{equation}
 Hence, in the stable and unstable vector basis $\left(\vec{\zeta_{u}},\vec{\zeta_{s}}\right)$,
the real matrices $\mathcal{\overline{C}}_{t}$ and $\mathcal{\overline{B}}_{t}$
take the form

\begin{equation}
\mathcal{\overline{C}}_{t}=\Re(\mathcal{\widetilde{V}})=\frac{1}{\left|\det\mathcal{V}_{t}\right|^{2}}\left[\mathit{C}\left(1+\tanh^{2}\left(t\lambda/2\right)\right)-2\mathit{B}_{t}\vec{\zeta_{u}}.\vec{\zeta_{s}}\tanh\left(t\lambda/2\right)\right],\label{eq:Cbar}
\end{equation}
and 

\begin{equation}
\mathcal{\overline{B}}_{t}=-\Im(\mathcal{\widetilde{V}})=\frac{1}{\left|\det\mathcal{V}_{t}\right|^{2}}\left[\mathit{B}_{t}\left(1+\tanh^{2}\left(t\lambda/2\right)\right)+2\mathit{C}\vec{\zeta_{u}}.\vec{\zeta_{s}}\tanh\left(t\lambda/2\right)\right]\label{eq:Bbar}
\end{equation}
with the symmetric matrix $\mathit{B}_{t}$, the scalar product matrix
$\mathit{C}$ and the determinant $\det\mathcal{V}_{t}$ respectively
given by the expressions (\ref{cayley}), (\ref{X-xr}) and (\ref{eq:moddetv}).
Inserting the expressions (\ref{eq:M+1}), (\ref{eq:Cbar}) and (\ref{eq:Bbar})
in the definition of the symmetric matrices $D_{t}$ and $E_{t}$
(\ref{eq:DyE}), we get 
\begin{equation}
D_{t}=-2\frac{\tanh\left(t\lambda/2\right)}{det_{1}}\left[\begin{array}{cc}-\zeta_{s}^{2}\left(\vec{\zeta_{u}}.\vec{\zeta_{s}}\right)e^{-t\lambda} & 1+\tanh^{2}\left(t\lambda/2\right)+2\left(\vec{\zeta_{u}}.\vec{\zeta_{s}}\right)^{2}\\
1+\tanh^{2}\left(t\lambda/2\right)+2\left(\vec{\zeta_{u}}.\vec{\zeta_{s}}\right)^{2} & -\zeta_{u}^{2}\left(\vec{\zeta_{u}}.\vec{\zeta_{s}}\right)e^{t\lambda}
\end{array}\right]\label{eq:Dlanda}
\end{equation}
and

\begin{equation}
E_{t}=-\frac{1+\tanh^{2}\left(t\lambda/2\right)}{det_{1}}\left[\begin{array}{cc}-\zeta_{s}^{2}e^{-t\nicefrac{\lambda}{2}} & \frac{\vec{\zeta_{u}}.\vec{\zeta_{s}}}{2\sinh^{2}\left(t\lambda/2\right)+1}\\
\frac{\vec{\zeta_{u}}.\vec{\zeta_{s}}}{2\sinh^{2}\left(t\lambda/2\right)+1} & -\zeta_{u}^{2}e^{t\nicefrac{\lambda}{2}}
\end{array}\right]\label{eq:Elanda}
\end{equation}
where we have defined 
\[
det_{1}=4\cosh^{2}\left(t\lambda/2\right)\left|\det\mathcal{V}_{t}\right|^{2}.
\]
It is important to note that, (\ref{eq:Dlanda}), (\ref{eq:Elanda}),
(\ref{cayley}), (\ref{eq:detv}) and (\ref{detM}) are respectively
explicit expression of the symmetric matrices $D_{t}$, $E_{t}$ and
$\mathit{B}_{t}$ and the determinants $\left|\det\mathcal{V}_{t}\right|$and
$\left|\det\left(\mathcal{M}_{\gamma}^{t}+1\right)\right|$ for any
value of the time $t$. Inserting these expressions in (\ref{eq:phiUphiSCcont}),
the time summation can be numerically performed. In this way, we obtain
a semiclassical expression for the matrix elements of the propagator
in the scar functions basis entirely in terms of classical features
such as, the chord $\xi_{0}$ that joins the points $X_{2}$ and $X_{1}$,
the action of the periodic orbit $\widetilde{S}_{X_{2}}$, the stable
and unstable vectors $\vec{\zeta_{u}},\vec{\zeta_{s}}$ and the Lyapunov
exponent $\lambda$.

\section{Scar functions Matrix elements for the Cat Map}

Now the present theory is applied to the cat map i.e. the linear automorphism
on the $2$-torus generated by the $2\times2$ symplectic matrix $\mathcal{M}$,
that takes a point $x_{-}$ to a point $x_{+}$ : $x_{+}=\mathcal{M}x_{-}\quad\mbox{mod(1)}$.
In other words, there exists an integer $2$-dimensional vector $\mathbf{m}$
such that $x_{+}=\mathcal{M}x_{-}-\mathbf{m}$. Equivalently, the
map can also be studied in terms of the center generating function
\cite{mcat}. This is defined in terms of center points 
\begin{equation}
x\equiv\frac{x_{+}+x_{-}}{2}\label{xdef}
\end{equation}
 and chords 
\begin{equation}
\xi\equiv x_{+}-x_{-}=-\mathcal{J}\frac{\partial S(x,\mathbf{m})}{\partial x},\label{cordef}
\end{equation}
 where 
\begin{align}
S(x,\mathbf{m}) & =xBx+x\left(B-\mathcal{J}\right)\mathbf{m}+\frac{1}{4}\mathbf{m}(B+\widetilde{\mathcal{J}})\mathbf{m}\label{sx2}
\end{align}
 is the center generating function. Here $B$ is a symmetric matrix
(the Cayley parameterization of $\mathcal{M}$, as in (\ref{cayley})),
while 
\begin{equation}
\widetilde{\mathcal{J}}=\left[\begin{array}{c|c}
0 & 1\\
\hline 1 & 0
\end{array}\right].
\end{equation}
 We will study here the cat map with the symplectic matrix 
\begin{equation}
\mathcal{M}=\left[\begin{array}{cc}
2 & 3\\
1 & 2
\end{array}\right]\mbox{, and symmetric matrix }B=\left[\begin{array}{cc}
-\frac{1}{3} & 0\\
0 & 1
\end{array}\right].\label{mhb}
\end{equation}
 This map is known to be chaotic, (ergodic and mixing) as all its
periodic orbits are hyperbolic. The periodic points $x_{l}$ of integer
period $l$ are labeled by the winding numbers $\mathbf{m,}$ so that
\begin{equation}
x_{l}=\left(\begin{array}{l}
{p_{l}}\\
{q_{l}}
\end{array}\right)=(\mathcal{M}^{l}-1)^{-1}\mathbf{m}.\label{xfix}
\end{equation}
 The first periodic points of the map are the fixed points at $(0,0)$
and $(\frac{1}{2},\frac{1}{2})$ and the periodic orbits of period
2 are $[(0,\frac{1}{2})$ , $(\frac{1}{2},0)]$, $[(\frac{1}{2},\frac{1}{6})$
, $(\frac{1}{2},\frac{5}{6})]$, $[(0,\frac{1}{6})$ , $(\frac{1}{2},\frac{2}{6})]$,
$[(0,\frac{5}{6})$, $(\frac{1}{2},\frac{4}{6})]$ and $[(0,\frac{2}{6}),(0,\frac{4}{6})]$.
The eigenvalues of $\mathcal{M}$ are $e^{-\lambda}$ and $e^{\lambda}$
with $\lambda=\ln(2+\sqrt{3})\approx1.317$. This is then the stability
exponent for the fixed points, whereas the exponents must be doubled
for orbits of period 2. All the eigenvectors have directions $\vec{\zeta_{s}}=(-\frac{\sqrt{3}}{2},\frac{1}{2})$
and $\vec{\zeta_{u}}=(1,\frac{1}{\sqrt{3}})$ corresponding to the
stable and unstable directions respectively.

Quantum mechanics on the torus, implies a finite Hilbert space of
dimension $N=\frac{1}{2\pi\hbar}$, and that positions and momenta
are defined to have discrete values in a lattice of separation $\frac{1}{N}$
\cite{hanay,opetor}. The cat map was originally quantized by Hannay
and Berry \cite{hanay} in the coordinate representation the propagator
is: 
\begin{equation}
\langle\mathbf{q}_{k}|\hat{\mathbf{U}}_{\mathcal{M}}|\mathbf{q}_{j}\rangle=\left(\frac{i}{N}\right)^{\frac{1}{2}}{\exp}\left[\frac{i2\pi}{N}(k^{2}-jk+j^{2})\right],\label{uqq}
\end{equation}
 where the states $\langle q|\mathbf{q}_{j}\rangle$ are periodic
combs of Dirac delta distributions at positions $q=j/Nmod(1)$, with
$j$ integer in $[0,N-1]$. In the Weyl representation \cite{opetor},
the quantum map has been obtained in \cite{mcat} as 
\begin{align}
\mathbf{U}_{\mathcal{M}}(x) & =\frac{2}{\left|\det(\mathcal{M}+1)\right|^{\frac{1}{2}}}\sum_{\mathbf{m}}e^{i2\pi N\left[S(x,\mathbf{m})\right]}\\
 & =\frac{2}{\left|\det(\mathcal{M}+1)\right|^{\frac{1}{2}}}\sum_{\mathbf{m}}e^{i2\pi N\left[xBx+x(B-\mathcal{J})\mathbf{m}+\frac{1}{4}\mathbf{m}(B+\widetilde{\mathcal{J}})\mathbf{m}\right]},\label{ugxp}
\end{align}
 where the center points are represented by $x=(\frac{a}{N},\frac{b}{N})$
with $a$ and $b$ integer numbers in $[0,N-1]$ for odd values of
$N$ \cite{opetor}. There exists an alternative definition of the
torus Wigner function which also holds for even $N$. 

The fact that the symplectic matrix $\mathcal{M}$ has equal diagonal
elements implies in the time reversal symmetry and then the symmetric
matrix $B$ has no off-diagonal elements. This property will be valid
for all the powers of the map and, using (\ref{ugxp}), we can see
that it implies in the quantum symmetry 
\begin{equation}
\mathbf{U}_{\mathcal{M}}^{l}(p,q)=\left(\mathbf{U}_{\mathcal{M}}^{l}(-p,q)\right)^{*}=\left(\mathbf{U}_{\mathcal{M}}^{l}(p,-q)\right)^{*}.\label{qsym}
\end{equation}
 for any integer value of $l$.

It has been shown \cite{hanay} that the unitary propagator is periodic
(nilpotent) in the sense that, for any value of $N$ there is an integer
$k(N)$ such that 
\[
\hat{\mathbf{U}}_{\mathcal{M}}^{k(N)}=e^{i\phi}.
\]
 Hence the eigenvalues of the map lie on the $k(N)$ possible sites
\begin{equation}
\left\{ \exp\left[\frac{i(2m\pi+\phi)}{k(N)}\right]\right\} ,\quad1\le m\le k(N).
\end{equation}
 For the cases where $k(N)\langle N$ there are degeneracies and the
spectrum does not behave as expected for chaotic quantum systems.
In spite of the peculiarities in this map, a very weak nonlinear perturbations
of cat maps restores the universal behavior of non degenerate chaotic
quantum systems spectra \cite{matos}. Eckhardt \cite{Eckhardt} has
argued that typically the eigenfunctions of cat maps are random.

The Scar Wigner Function on the torus depends on the definition of
the periodic coherent state \cite{nonen}, with $\langle p\rangle=P$
and $\langle q\rangle=Q$. In accordance to (\ref{eq:qX}) 
\begin{equation}
\langle{\bf X}|\mathbf{q}_{k}\rangle=\sum_{j=-\infty}^{\infty}\exp\left\{ -\frac{1}{\hbar}\left[iP(j+\frac{Q}{2}-\nicefrac{k}{N})+\frac{1}{2}(j+Q-\nicefrac{k}{N})^{2}\right]\right\} .\label{35}
\end{equation}
 The Scar function is then defined on the torus as 
\begin{equation}
\left|{\bf \varphi_{X,\phi}}\right\rangle =\sum_{t=-\infty}^{\infty}e^{i\phi t}e^{-\left(\frac{4t}{T}\right)^{2}}\mathbf{U}_{\mathcal{M}}^{t}\left|{\bf X}\right\rangle .\label{scartoro}
\end{equation}
 Remember that for maps, time only takes discrete values, then the
time integral in (\ref{scarfun}) has been in this case replaced by
a summation. Also, as we have already discuss, for our numerical computations
we truncate the sum for times $|t|>T/2$ where the Gaussian damping
term became negligible. 

In order to construct operators or functions on the torus we have
to periodize the construction. This is done merely using the recipe
\cite{opetor} that for any operator its Weyl representation on the
torus ${\bf A}(x)$ is obtained from is analogue in the plane $A(x)$
by 
\begin{align*}
{\bf A}(x)=\sum_{j=-\infty}^{\infty}\sum_{k=-\infty}^{\infty}(-1)^{2ja+2kb+jkN}A(x+\frac{(k,j)}{2}).
\end{align*}
Indeed the construction on the torus from the plane is obtain in terms
of averages over equivalent points, that are obtained by translation
with integer chords: $\widehat{T}_{\overrightarrow{k}}$ where $\overrightarrow{k}=\left(k_{p},k_{q}\right)$
is a two dimensional vector with integer components $k_{p}$ and $k_{q}$.
Hence, the unit operator in the Hilbert space of the torus is \cite{opetor}

\[
\boldsymbol{\widehat{1}_{N}}=\sum_{k=0}^{N-1}|\mathbf{q}_{k}\rangle\langle\mathbf{q}_{k}|=\biggl\langle\widehat{T}_{\overrightarrow{k}}e^{i2\pi(\chi\wedge\overrightarrow{k}+\frac{N}{4}\overrightarrow{k}{\mathcal{\widetilde{J}}}\overrightarrow{k})}\biggr\rangle
\]
 so that 
\[
|{\bf X}\rangle=\boldsymbol{\widehat{1}_{N}}|X\rangle=\biggl\langle e^{i2\pi(\chi\wedge\overrightarrow{k}+\frac{N}{4}\overrightarrow{k}{\mathcal{\widetilde{J}}}\overrightarrow{k})}\widehat{T}_{\overrightarrow{k}}|X\rangle\biggr\rangle=\biggl\langle e^{i2\pi(\chi\wedge\overrightarrow{k}+\frac{N}{4}\overrightarrow{k}{\mathcal{\widetilde{J}}}\overrightarrow{k})}e^{-\frac{i}{2\hbar}X\wedge\overrightarrow{k}}|X+\overrightarrow{k}\rangle\biggr\rangle
\]
 In this way the coherent sates matrix elements for any operator on
the torus are obtained through

\begin{equation}
\langle\mathbf{X}_{1}|\mathbf{\widehat{A}}|\mathbf{X}_{2}\rangle=\biggl\langle e^{i2\pi(\chi\wedge\overrightarrow{k}+\frac{N}{4}\overrightarrow{k}{\mathcal{\widetilde{J}}}\overrightarrow{k})}e^{-\frac{i}{2\hbar}X\wedge\overrightarrow{k}}\langle X_{1}|\widehat{A}|X_{2}+\overrightarrow{k}\rangle\biggr\rangle\label{eq:Atoro}
\end{equation}
 and analogously for the scar functions matrix elements.

In table 1 we compare, for different values of $N$ correspondingly
$\hbar=1/(2\pi N)$, the exact Scar matrix elements for a cat map
with the semi classical ones obtained with expression (\ref{eq:phiUphiSCcont})
taking in both cases the torus periodization (\ref{eq:Atoro}). 
As we can observe the semi classical expression (\ref{eq:phiUphiSCcont}) is exact in
this case.
This fact is not surprising since the cat map is equivalent to a quadratic Hamiltonian system.
Also, we have verified that, as was previously seen in \cite{13ver}, this matrix elements are no null
only for values of $N$ that are multiple of four and in this cases the matrix elements 
$\left\langle \varphi_{X_{1}}^{\phi_{1}}\right|\left.\varphi_{X_{2}}^{\phi_{2}}\right\rangle $ 
are real numbers. 

\begin{table}[H]
\begin{tabular}{|c|c|c|c|c|}
\hline 
$N$ & $\left\langle \varphi_{X_{1}}^{\phi_{1}}\right|\left.\varphi_{X_{2}}^{\phi_{2}}\right\rangle $ & $\left\langle \varphi_{X_{1}}^{\phi_{1}}\right|\left.\varphi_{X_{2}}^{\phi_{2}}\right\rangle _{SC}$ & $\left\langle \varphi_{X_{1}}^{\phi_{1}}\right|\widehat{U}\left|\varphi_{X_{2}}^{\phi_{2}}\right\rangle $ & $\left\langle \varphi_{X_{1}}^{\phi_{1}}\right|\widehat{U}\left|\varphi_{X_{2}}^{\phi_{2}}\right\rangle _{SC}$\tabularnewline
\hline 
\hline 
100 & $0.32170130 $ & $0.32170128 $ & $ 0.36448490 + i 0.419063829 $ & $ 0.36448489 + i 0.419063830 $\tabularnewline
\hline 
101 & $0,0$ & $0,0$ & $0,0$ & $0,0$\tabularnewline
\hline 
104 & $0.33082419 $ & $0.33082418 $ & $ 0.37520132 + i 0.397240638 $ & $ 0.37520133 + i 0.397240637 $\tabularnewline
\hline 
200 & $0.36468529 $ & $0.36468530 $ & $ 0.45326651 + i 0.358242569 $ & $ 0.45326650 + i 0.358242570 $\tabularnewline
\hline 
\end{tabular}

\caption{Numerical comparison between the exact scar functions matrix elements
and scar functions matrix elements obtained from expression (\ref{eq:phiUphiSCcont}).
First column shows the different values of $N=1/(2\pi\hbar)$, the
inverse of the Plank constant. Second column displays (real and imaginary
parts) the exact of the matrix elements between the two scar functions
constructed on the fixed points of the cat map. Third column shows
the respective semi classical approximation, using expression (\ref{eq:phiUphiSCcont}),
of the matrix elements shown in the second column. Fourth column displays
(real and imaginary parts) the exact matrix elements between the two
scar functions constructed on the fixed points now for one iteration
of the quantum propagator for the cat map. Fifth column shows the
respective semi classical approximation, using expression (\ref{eq:phiUphiSCcont}),
of the matrix elements shown in the forth column.}

\end{table}

\section{Discussion}

The semiclassical theory of short periodic orbits developed by Vergini
and co-workers \cite{6ver}-\cite{13ver} is a formalism where the
number of used periodic orbits needed to obtain the spectrum of a
classically chaotic system increases only linearly with the mean energy
density, allowing to obtain all the quantum information of a chaotic
Hamiltonian system in terms of a very small number of short periodic
orbits. The key elements in this theory are wave functions related
to short unstable POs and then, it is crucial the evaluation of matrix
elements between these wave functions.

In this work by means of the Weyl representation, we have obtained
a semiclassical expression for this matrix elements of the propagator
in scar functions basis entirely in terms of the classical canonical invariants
such as, the chord that joins the points $X_{2}$ and $X_{1}$, the
action of the periodic orbit, the stable and unstable vectors $\vec{\zeta_{u}},\vec{\zeta_{s}}$
and the Lyapunov exponent $\lambda$. Also, the comparison with a
system whose semiclassical limit is exact has allowed to correctly
check the exactness of the obtained expression up to quadratic Hamiltonian
systems. 

As has been already seen \cite{13ver,epldiag}, with these matrix
elements at hand, the spectrum of the propagator can be obtained without
requiring an explicit computation of scar functions. Of course, in
order to include nonlinear contributions a deeper understanding of
the dynamics up to the Ehrenfest time is required; in this respect,
enormous efforts were recently carried out in such a direction for
the diagonal matrix elements \cite{epldiag}.

\section*{Acknowledgments}

I am grateful to E. Vergini, M. Saraceno and G. Carlo for stimulating
discussions and thanks the CONICET for financial support.

\appendix
\section*{Appendix: Reflection Operators in Phase Space}

Among the several representations of quantum mechanics, the Weyl-Wigner
representation is the one that performs a decomposition of the operators
that acts on the Hilbert space, on the basis formed by the set of
unitary reflection operators. In this appendix we review the definition
and some properties of this reflection operators.

First of all we construct the family of unitary operators 
\begin{equation}
\hat{T}_{q}=\exp(-i\hbar^{-1}q.\hat{p}),\qquad\hat{T}_{p}=\exp(i\hbar^{-1}p.\hat{q}),
\end{equation}
 and following \cite{ozrep}, we define the operator corresponding
to a general translation in phase space by $\xi=(p,q)$ as 
\begin{align}
  \hat{T}_{\xi} & \equiv &\exp\left(\frac{i}{\hbar}\xi\wedge\hat{x}\right)\equiv\exp\left[\frac{i}{\hbar}(p.\hat{q}-q.\hat{p})\right]\\
 & = & \hat{T}_{p}\hat{T}_{q}\ \exp\left[-\frac{i}{2\hbar}\ p.q\right]=\hat{T}_{q}\hat{T}_{p}\ \exp\left[\frac{i}{2\hbar}\ p.q\right]\ ,\label{eq:tcor}
\end{align}
 where naturally $\hat{x}=(\hat{p},\hat{q})$. In other words, the
order of $\hat{T}_{p}$ and $\hat{T}_{q}$ affects only the overall
phase of the product, allowing us to define the translation as above.
$\hat{T}_{\xi}$ is also known as a \textit{Heisenberg operator}.
Acting on the Hilbert space we have: 
\begin{equation}
\widehat{T}_{\xi}|q_{a}\rangle=e^{\frac{i}{\hbar}p(q_{a}+\frac{q}{2})}|q_{a}+q\rangle\label{eq:tq}
\end{equation}
 and 
\begin{equation}
\widehat{T}_{\xi}|p_{a}\rangle=e^{-\frac{i}{\hbar}q(p_{a}+\frac{p}{2})}|p_{a}+p\rangle.
\end{equation}
 We, hence, verify their interpretation as translation operators in
phase space. The group property is maintained within a phase factor:
\begin{equation}
\hat{T}_{\xi_{2}}\hat{T}_{\xi_{1}}=\hat{T}_{\xi_{1}+\xi_{2}}\ \exp[\frac{-i}{2\hbar}\xi_{1}\wedge\xi_{2}]=\hat{T}_{\xi_{1}+\xi_{2}}\ \exp[\frac{-i}{\hbar}D_{3}(\xi_{1},\xi_{2})],\label{eq:tt}
\end{equation}
 where $D_{3}$ is the symplectic area of the triangle determined
by two of its sides. Evidently, the inverse of the unitary operator
$\hat{T}_{\xi}^{-1}=\hat{T}_{\xi}^{\dag}=\hat{T}_{-\xi}$ .

The set of operators corresponding to phase space reflections $\hat{R}_{x}$
about points $x=(p,q)$ in phase space, is formally defined in \cite{ozrep}
as the Fourier transform of the translation (or Heisenberg) operators
\begin{equation}
\widehat{R}_{x}\equiv(4\pi\hbar)^{-L}\int d\xi\quad e^{\frac{i}{\hbar}x\wedge\xi}\widehat{T}_{\xi}.\label{eq:rint}
\end{equation}
 Their action on the coordinate and momentum bases are 
\begin{align}
\hat{R}_{x}\left|q_{a}\right\rangle  & = & e^{2i(q-q_{a})p/\hbar}\;\left|2q-q_{a}\right\rangle \label{rqpl}\\
\hat{R}_{x}\left|p_{a}\right\rangle  & = & e^{2i(p-p_{a})q/\hbar}\;\left|2p-p_{a}\right\rangle ,
\end{align}
 displaying the interpretation of these operators as reflections in
phase space. Also, Using the coordinate representation of the coherent
state (\ref{eq:qX}) and the action of reflection on the coordinate
basis (\ref{rqpl}), we can see that the action of the reflection
operator $\widehat{R}_{x}$ on a coherent state $\left|X\right\rangle $
is the $x$ reflected coherent state 
\begin{equation}
\widehat{R}_{x}\left|X\right\rangle =\exp\left(\frac{i}{\hbar}X\wedge x\right)\left|2x-X\right\rangle .
\end{equation}

This family of operators have the property that they are a decomposition
of the unity (completeness relation) 
\begin{equation}
\hat{1}=\frac{1}{2\pi\hbar}\int dx\ \hat{R}_{x},\label{1R}
\end{equation}
 and also they are orthogonal in the sense that 
\begin{equation}
Tr\left[\hat{R}_{x_{1}}\ \hat{R}_{x_{2}}\right]=2\pi\hbar\;\delta(x_{2}-x_{1}).\label{trR}
\end{equation}
 Hence, an operator $\hat{A}$ can be decomposed in terms of reflection
operators as follows 
\begin{equation}
\hat{A}=\frac{1}{2\pi\hbar}\int dx\ A_{W}(x)\ \hat{R}_{x}.\label{rep}
\end{equation}
 With this decomposition, the operator $\hat{A}$ is mapped on a function
$A_{W}(x)$ living in phase space, the so called Weyl-Wigner symbol
of the operator. Using (\ref{trR}) it is easy to show that $A_{W}(x)$
can be obtained by performing the following trace operation 
\[
A_{W}(x)=Tr\left[\hat{R}_{x}\ \hat{A}\right].
\]
 Of course, as it is shown in \cite{ozrep}, the Weyl symbol also
takes the usual expression in terms of matrix elements of $\hat{A}$
in coordinate representation 
\[
A_{W}(x)=\int\left\langle q-\frac{Q}{2}\right|\hat{A}\left|q+\frac{Q}{2}\right\rangle \exp\left(-\frac{i}{\hbar}pQ\right)dQ.
\]

It was also shown in \cite{ozrep} that reflection and translation
operators have the following composition properties 
\begin{equation}
\widehat{R}_{x}\widehat{T}_{\xi}=\widehat{R}_{x-\xi/2}e^{-\frac{i}{\hbar}x\wedge\xi}\ ,\label{eq:rt}
\end{equation}
 
\begin{equation}
\widehat{T}_{\xi}\widehat{R}_{x}=\widehat{R}_{x+\xi/2}e^{-\frac{i}{\hbar}x\wedge\xi}\ ,\label{eq:tr}
\end{equation}
 
\begin{equation}
\widehat{R}_{x_{1}}\widehat{R}_{x_{2}}=\widehat{T}_{2(x_{2}-x_{1})}e^{\frac{i}{\hbar}2x_{1}\wedge x_{2}}\label{eq:rr}
\end{equation}
 so that 
\begin{equation}
\widehat{R}_{x}\widehat{R}_{x}=\widehat{1}\ .
\end{equation}
 Now using (\ref{eq:rr}) and (\ref{eq:tr}) we can compose three
reflections so that 
\begin{equation}
\widehat{R}_{x_{2}}\widehat{R}_{x}\widehat{R}_{x_{1}}=e^{\frac{i}{\hbar}\Delta_{3}(x_{2},x_{1},x)}\widehat{R}_{x_{2}-x+x_{1}}
\end{equation}
 where $\Delta_{3}(x_{2},x_{1},x)=2(x_{2}-x)\wedge(x_{1}-x)$ is the
area of the oriented triangle whose sides are centered on the points
$x_{2},x_{1}$ and $x$ respectively.

\end{document}